\documentclass[aps,prl,twocolumn,superscriptaddress]{revtex4}
\usepackage{mathrsfs}
\usepackage{epsfig}
\usepackage{graphicx}
\usepackage{amsfonts}
\usepackage[figuresright]{rotating}
\usepackage{amssymb}
\usepackage{amsmath}
\usepackage{dcolumn}
\usepackage{bm}
\usepackage{color}

\def\be{\begin{equation}} \def\ee{\end{equation}}
\def\bea{\begin{eqnarray}} \def\eea{\end{eqnarray}}

\newcommand{\WQCASQC} { Wilczek Quantum Center and Shanghai Research Center for Quantum Sciences, School of Physics and Astronomy, Shanghai Jiao Tong University, Shanghai 200240, China}

\newcommand{\Key}{Key Laboratory of Artificial Structures and Quantum Control, School of Physics and Astronomy, Shanghai Jiao Tong University, Shanghai 200240, China}

\newcommand{\Hefei} {Hefei National Laboratory, Hefei 230088, China}

\begin{document}
\title{Floquet-induced bosonic pair condensate with unconventional symmetry}

\author{Zhizhen Chen}
\thanks{These authors contributed equally to this work}
\affiliation{\WQCASQC}

\author{Jiale Huang}
\thanks{These authors contributed equally to this work}
\affiliation{\Key}

\author{Mingpu Qin}
\email{qinmingpu@sjtu.edu.cn}
\affiliation{\Key}
\affiliation{\Hefei}

\author{Zi Cai}
\email{zcai@sjtu.edu.cn}
\affiliation{\WQCASQC}
\affiliation{\Key}

\begin{abstract}  
In this study, we propose a dynamical pairing mechanism other than the pair-wise interactions. Starting from a two-dimensional hard-core boson model with periodically modulated hopping amplitude, we derive an effective Floquet Hamiltonian with three-site interactions that are responsible for unconventional pairing between adjacent bosons. By performing a density matrix renormalization group 
study on this three-site interacting Hamiltonian, we reveal a  bosonic pair condensate with $s+id$ wave symmetry, while the single-particle Bose-Einstein condensate is completely depleted.  The experimental implementations of the proposed model on superconducting quantum circuit have also been discussed.
\end{abstract}


\maketitle

{\it Introduction --}  Searching for macroscopic quantum coherent states with unconventional pairing and exploring their pairing mechanism have been focus across several disciplines of quantum physics in the past decades. Examples in this regards include the d-wave pairing in cuprate superconductors\cite{Anderson1987,Gros1988,Kotliar1988,Tsuei2000} and cold atoms\cite{Trebst2006,Mark2025} and p-wave pairing in $^3$He\cite{Volovik2003}, fractional quantum hall\cite{Moore1991} and topological superconductor\cite{Read2000}, while most studies in this regards focus on equilibrium systems wherein the pairing is induced by conservative forces between particles. However, for a non-equilibrium quantum system, the existence of the unconventional pairing states and the mechanism behind them remain elusive\cite{Diehl2010b,Foster2014a}.  Recently, the experimental progresses in synthetic quantum systems including cold atoms, Rydberg atomic array and superconducting quantum circuit have revealed unprecedented opportunities for exploring the non-equilibrium quantum many-body states\cite{Schreiber2015,Zhang2017,Choi2017,Bernien2017,Kyprianidis2021,Mi2022,Bluvstein2023}, thus one may wonder whether it is possible to realize nonequilibrium unconventional pairing states in these systems, if so, how to distinguish them from  their equilibrium counterparts?

In past decades, the periodic driving has been widely used as a knot to manipulate the properties of quantum many-body systems and realize the quantum  matters inaccessible in conventional equilibrium systems\cite{Oka2009,Lindner2011,Zhou2023,Struck2011,Jotzu2014}. Motivated by these progresses, we propose a driving protocol wherein a simple periodic modulation of single-particle hopping suffices to result in unconventional pairing condensate.  In stark contrast to the pairwise interactions in equilibrium condensate, this unconventional pairing originates from an emergent three-site interaction that is uniquely tied to nonequilibrium feature of the driven system\cite{Choi2020,Petiziol2021,Wu2025}. Although three or higher-body interactions are not present at a fundamental level, they can appear in effective theories, and are responsible for a plethora of intriguing quantum states\cite{Moessner2001,Balents2002,Kitaev2003,Raussendorf2003,BUCHLER2007,Smacchia2011}.  In contrast to most equilibrium systems where the two-particle interaction dominates while the three-body interactions only provide small corrections due to their  perturbative character, in our non-equilibrium setup, the two-body terms in the effective Hamiltonian are completely suppressed, thus the three-body terms dominate and provide a pairing mechanism beyond the pair-wise interactions.


\begin{figure}[htbp]
	\centering
	\includegraphics[width=0.49\textwidth]{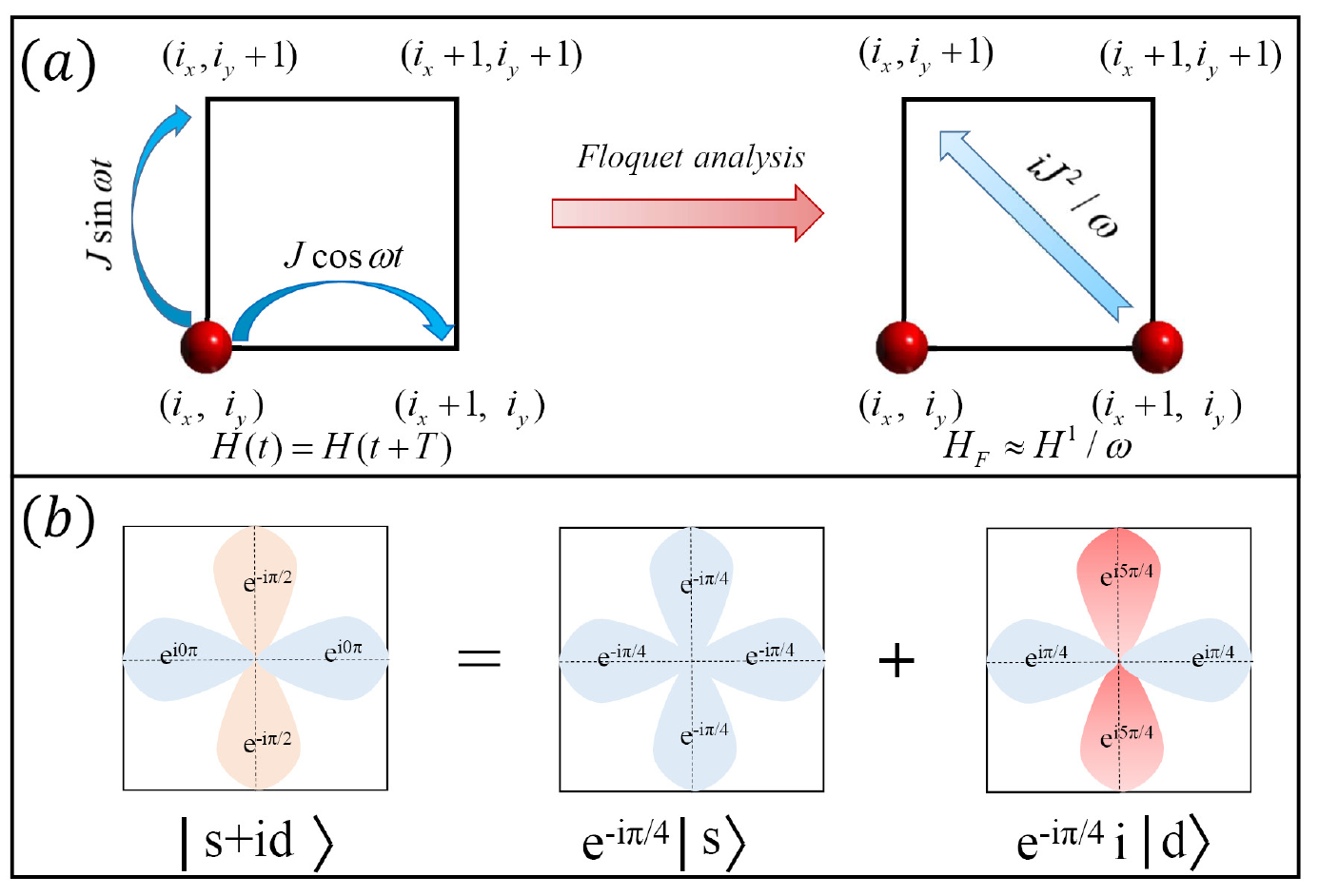}
	\caption{(a)Sketch of the original hard-core boson Hamiltonian with periodic modulation of the hopping amplitude and the time-independent Floquet Hamiltonian (at high frequency) with correlated hopping. (b) Phase distribution of the order parameter of the $s+id_{x^2-y^2}$ pair condensate}
	\label{fig:fig1}
\end{figure}

In this study, we propose a two-dimensional (2D) hard-core boson model, where the single-particle hopping amplitudes along the horizontal and vertical directions are periodically modulated in the same frequency but with a $\pi/2$ phase lag, and no other interactions than the hard-core constraint are present. In the presence of fast driving, it is shown that the stroboscopic dynamics of such a periodically driven system is governed by a time-independent Floquet Hamiltonian, where the leading terms involve three-site interactions that favor pairing between hard-core bosons on adjacent sites. By perform a density matrix renormalization group (DMRG) study\cite{White1992,Schollwock2005} on such an effective Floquet Hamiltonian, we reveal a  $s+i d_{x^2-y^2}$ wave bosonic pair condensate state, where the single-particle condensate is completely suppressed. The experimental realization of our model  has also been discussed.

\begin{figure}[htbp]
	\centering
	\includegraphics[width=0.49\textwidth]{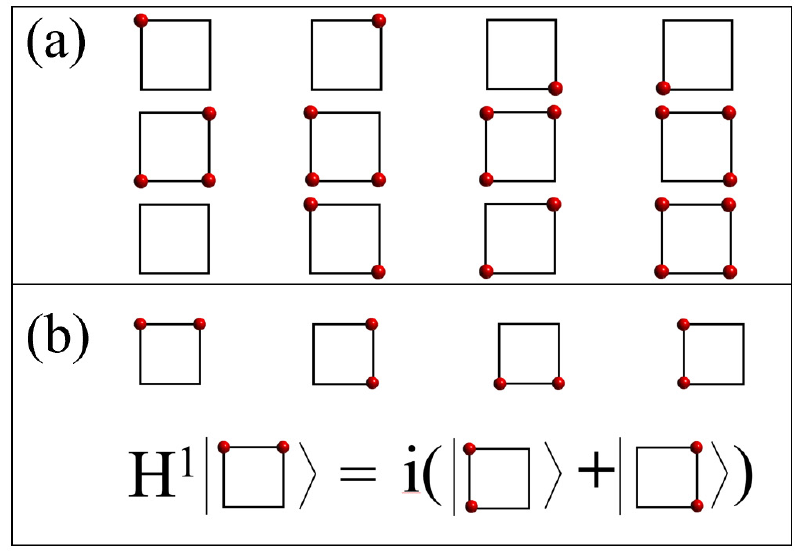}
	\caption{(a) Dark Fock states $|\sigma_D\rangle$ of $H^1$ in a single plaquette ($H^1|\sigma_D\rangle=0$).   (b) The 4 Fock states outside the dark state manifold in a single plaquette, each of which is connected to two others via $H^1$.}
	\label{fig:fig2}
\end{figure}

{\it Model and Floquet analysis --} The proposed model is a hard-core boson model in a 2D $L_x\times L_y$ square lattice with a time-dependent hopping amplitude, and the Hamiltonian reads:
\begin{eqnarray}
\nonumber H(t)=&J&\sum_{i_x,i_y} \{\cos\omega t [a_{i_x,i_y}^\dag a_{i_x+1,i_y}+a_{i_x+1,i_y}^\dag a_{i_x,i_y}]\\
&+&\sin\omega t [a_{i_x,i_y}^\dag a_{i_x,i_y+1}+a_{i_x,i_y+1}^\dag a_{i_x,i_y}]\} \label{eq:Ham}
\end{eqnarray}    
where $a_\mathbf{i}^\dag$ ($a_\mathbf{i}$) is the creation (annihilation) operator for the hard-core boson at site $\mathbf{i}=(i_x,i_y)$, which satisfies the commutation relation: $[a_{\mathbf{i}},a^\dag_{\mathbf{j}}]=\delta_{\mathbf{ij}} 2(n_\mathbf{i}-1)$ with $n_\mathbf{i}=a^\dag_{\mathbf{i}} a_{\mathbf{i}}$ being the density operator of the hard-core boson.  $J$ is the amplitude of the single-particle hopping between adjacent sites, $\omega$ is the frequency of the periodic driving ($\omega=2\pi/T$ with $T$ being the driving period).  

In general, the stroboscopic dynamics of a periodically driven system with $H(t)=H(t+T)$ can be described by a time-independent Floquet Hamiltonian $H_F$, which is defined as:
\begin{equation}
e^{-i T H_F}=\mathcal{T} e^{-i\int_0^T dt H(t)}
\end{equation}
where $\mathcal{T}$ is the chronological operator. If the driving is sufficiently fast $(\omega \gg J)$, the Floquet Hamiltonian that can be expressed in terms of the Magus expansion as:
\begin{equation}
	H_F=H^{0}+\frac 1\omega H^{1} +\frac 1{\omega^2} H^{2}+\dots \label{eq:Floquet}
\end{equation}
The zero-order term $H^{0}$ is a time-averaged Hamiltonian over a period: $H^{0}=\frac 1T \int_0^T dt H(t)$, {\it which is exactly zero in our setup}. Therefore, the dynamics is governed by the first order term $H^{1}$, which takes the form: 
\begin{equation}
H^{1}=\sum_{l=1}^\infty \frac 1l [H_l, H_{-l}], \label{eq:Ham1}
\end{equation}   
where $H_l$ is the $l$-th Fourier component of $H(t)$: $H(t)=\sum_{l=-\infty}^\infty e^{il\omega t} H_l$. Specific to our model with the Hamiltonian.(\ref{eq:Ham}), the $H_l$  terms with $l\neq \pm 1$ vanish. By substituting $H_{\pm 1}$ into Eq.(\ref{eq:Ham1}), one can derive $H_F$ up to the 1st order of $T$ as: 
\begin{small}
\begin{eqnarray}
H_F&=&\frac {1}\omega  H^{1}+\mathcal{O}\big(\frac{J^3}{\omega^2}\big ) \label{eq:Hamff}\\  \nonumber H^{1}&=&iJ^2\sum_{i_x,i_y}\big[(n_{i_x+1,i_y+1}-n_{i_x,i_y})(a^\dag_{i_x,i_y+1}a_{i_x+1,i_y}-h.c)\\
&+&(n_{i_x+1,i_y}-n_{i_x,i_y+1})(a^\dag_{i_x,i_y}a_{i_x+1,i_y+1}-h.c)\big]
 \label{eq:HamF}
\end{eqnarray}  
\end{small}
The leading term $H^{1}$ in the Floquet Hamiltonian is composed of the terms with a density operator coupled to a current operator along the diagonal bonds in each plaquette of the square lattice, as shown in Fig.\ref{fig:fig1} (a).  They resemble the density-assisted hopping (or correlated hopping) terms, which might be relevant to the high-Tc superconductor\cite{HIRSCH1989,Japaridze1999,Jiang2023,Lisandrini2025}, and have been directly observed in cold atom experiments\cite{Jurgensen2014,Meinert2016,Klemmer2024}. In both cases, the typical amplitude of the density-assisted hopping is much smaller than that of the bare hopping, in stark contrast to our model in which the bare hopping is completely suppressed.

By comparing the effective  Hamiltonian in Eq.(\ref{eq:HamF}) to its original Hamiltonian in Eq.(\ref{eq:Ham}), we can find that both of them preserve the U(1) symmetry, which corresponds to the total particle number conservation. In addition to that, it is worthy to mention that there is an emergent symmetry in Hamiltonian in Eq.(\ref{eq:HamF}), which corresponds to extra conserved quantities.  Notice that a bipartite lattice ({\it e.g.} square lattice) can be divided into two sublattices, and there is no particle hopping between them in $H^1$, therefore the total particle number in each sublattice is also conserved. However, this conservation only appears in $H^{1}$, thus is not exact and will be violated once higher order terms in the Magnus expansion are taken into account.   
\begin{figure}[htbp]
	\centering
	\includegraphics[width=0.49\textwidth]{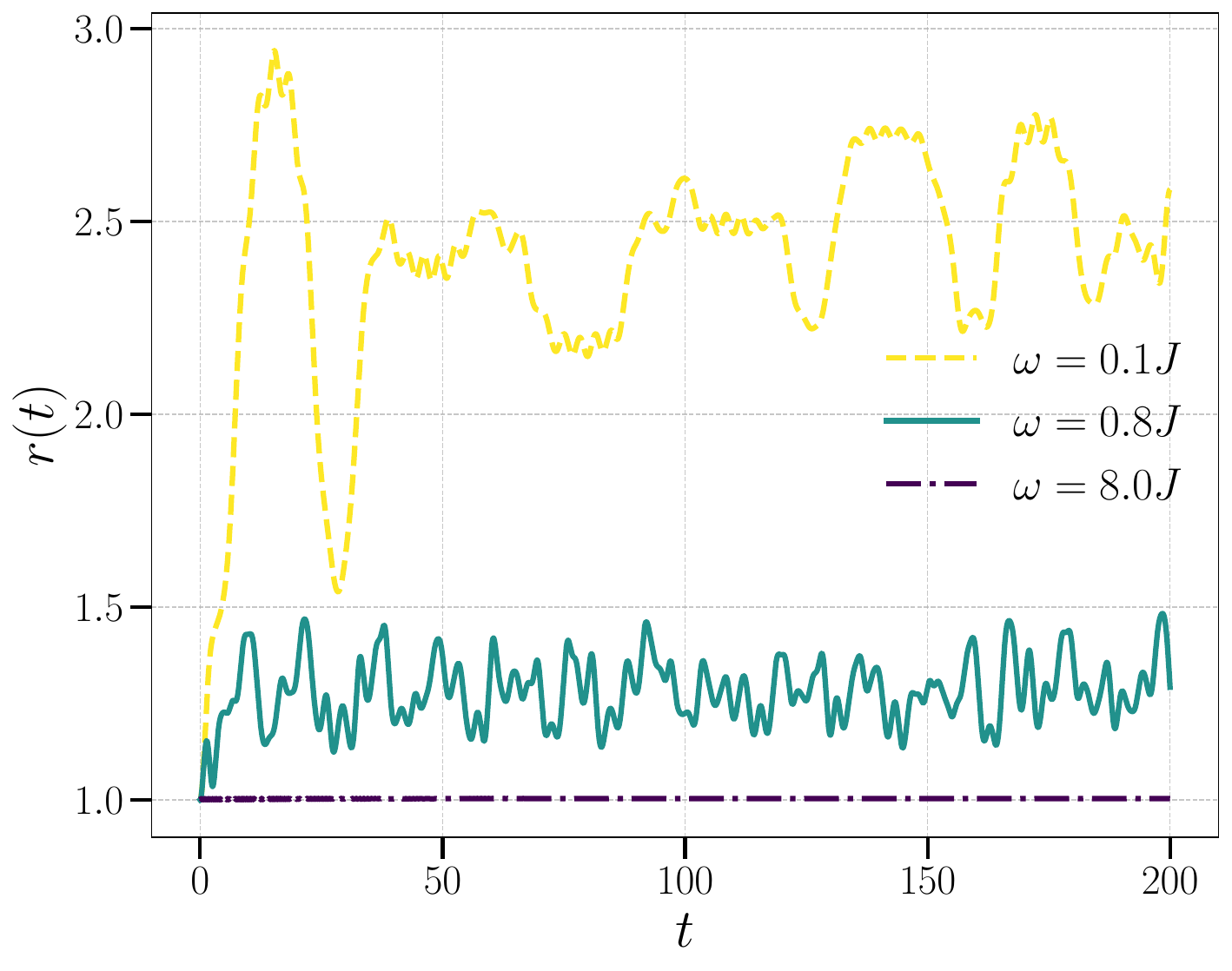}
	\caption{The evolution of the average distance between the two bosons $r(t)$ in a two-leg ladder ($12\times 2$) system  at different driving frequencies.}
	\label{fig:fig3}
\end{figure}

\begin{figure*}[htb]
	\includegraphics[width=1.0\textwidth]{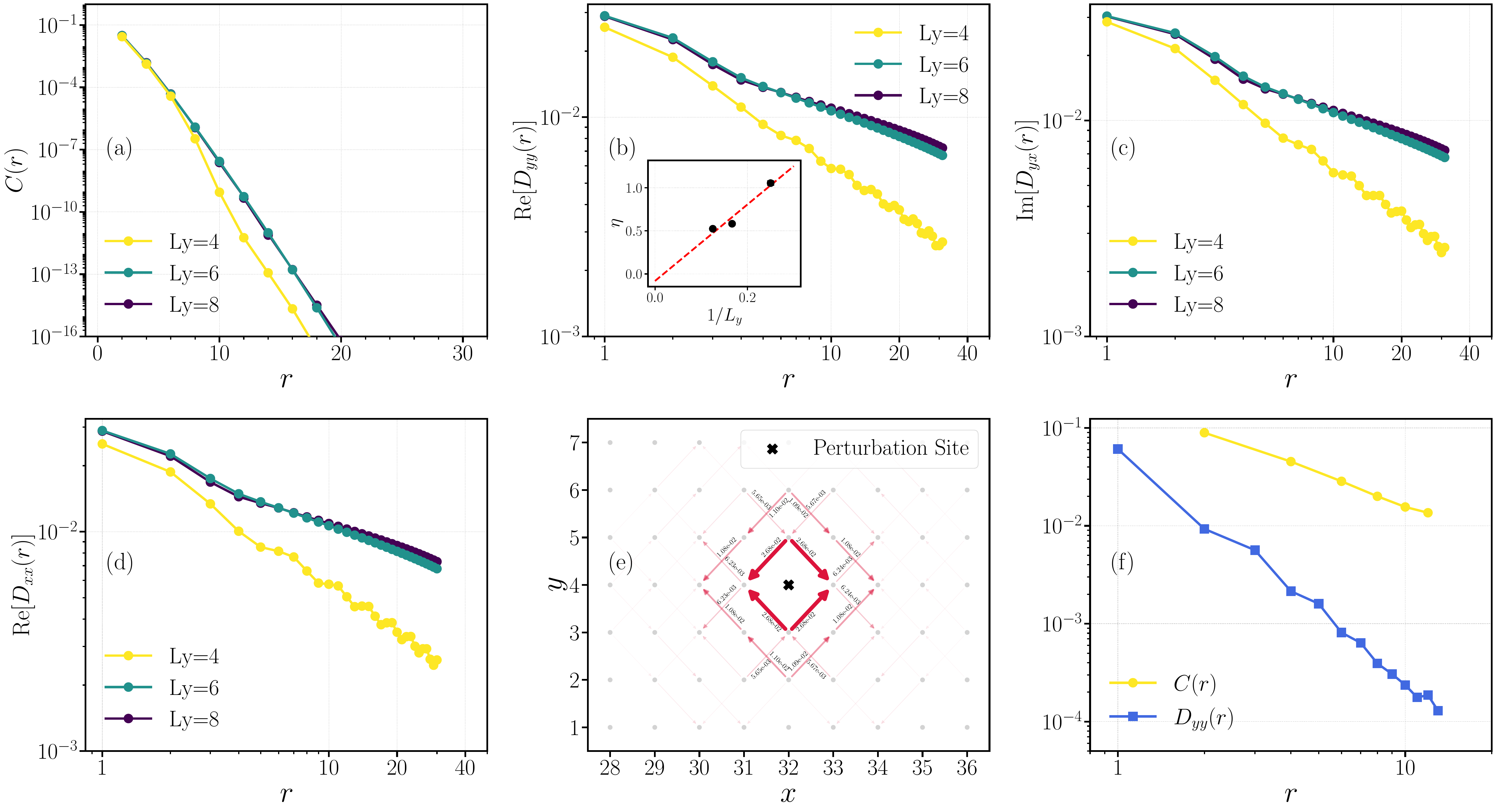}
	\caption{In the ground state of $H^1$, (a) the single particle correlation functions, and (b)-(d) various pair correlation functions defined in Eq.(\ref{eq:correlation}) in a 2D cylindrical lattice with a fixed $L_x=64$ and various $L_y$ at 1/8 filling. (b)-(d) exhibit power-law decays $\sim r^{-\eta}$, and the inset of (b) indicates the dependence  of the power exponent $\eta$ on $1/L_y$. (e) The spatial distribution of the currents along the diagonal bonds in the presence of a nonmagnetic impurity with local chemical potential $\mu_c=J^2/\omega $ on the origin point $(L_x/2,L_y/2)$ of the cylindrical lattice with a  $L_x=64$ and $L_y=8$ at 1/8 filling. (f) The single-particle and pair correlation functions in 2D cylindrical lattice with $L_x=32$ and $L_y=4$ at half filling.}
	\label{fig:fig4}
\end{figure*}

{\it Floquet induced dark states and pairing--} To gain some insight on the properties of the Floquet Hamiltonian $H^1$ at fast driving, we first consider a single plaquette with the $2^4=16$ Fock bases, while 12 of them (denoted by $|\sigma_D\rangle$ as shown in Fig.\ref{fig:fig2} (a))  are the ``dark states": these Fock states are annihilated by $H^1$ ($H^1|\sigma_D\rangle=0$). This fact enables us to construct a set of dark states of $H^1$ in a 2D lattice, whose  number  diverges exponentially with the system size\cite{Supplementary}. They barely evolve under $H_F$ in the presence of fast driving where $H_F\approx H_1/\omega$.

After eliminating the dark states, the Hilbert space of the system becomes highly constrained. Still taking a single plaquette for an example, as shown in Fig.\ref{fig:fig2} (b), each of the remaining 4 Fock states contains two bosons placed on a pair of adjacent sites, and is connected with two others via $H^1$.  This fact can also be generalized to a 2D lattice: if there are only two bosons in total and they are separated in distance, all the plaquettes are in the dark state, thus the bosons are stuck. On the contrary, if they are placed on a pair of adjacent sites, they can propagate in the lattice, but only move in pair from one bond to another.  

This picture can be numerically verified  by calculating  the dynamics of the system starting from an initial state with a pair of nearest-neighboring bosons, and evolving under the original periodic Hamiltonian $H(t)$ in Eq.(\ref{eq:Ham})  instead of $H^1$. Since the particle number is conserved, the wavefunction  at time t $|\psi(t)\rangle$  can be expanded in terms of the Fock basis ($|\sigma\rangle$) of the Hilbert space with 2 bosons only as  $|\psi(t)\rangle=\sum_\sigma c_\sigma(t)|\sigma\rangle$, where $c_\sigma(t)$ is the projection coefficient of $|\psi(t)\rangle$ on the basis $|\sigma\rangle$.   To monitor the relative distance between the bosons,  we define the average distance between the bosons as: $r(t)=\sum_\sigma |c_\alpha(t)|^2 r_\sigma$, where $r_\sigma$ is the distance between the two bosons in the $\sigma$ Fock basis. As shown in Fig.\ref{fig:fig3}, for a fast driving, $r(t)$ keeps its initial value and barely change in time ($r(t)\approx 1$), which agrees with the prediction of the Floquet analysis that the bosons can  hop only in pair under the evolution of $H^1$. For slower driving, $r(t)$ starts to oscillate, and will grow in time when the driving period is further increased, indicating that the two bosons are separated in space. The breakdown of pairing is due to the higher order terms other than $H^1$ in Eq.(\ref{eq:HamF}), whose effect cannot be neglected when the driving is sufficiently slow, thus make the system violate the kinetic constraint imposed by $H^1$.

{\it $s+i d_{x^2-y^2}$ pair condensate with no single-particle condensation --} In the following,  we will consider the many-body situation, where we restrict our discussion on the ground state of $H^1$, which corresponds to one of the eigenstate of the Floquet Hamiltonian $H_F$ in the fast driving limit (there is no ``ground state" for $H_F$ due to the periodicity of its energy spectrum). Such a state, once prepared, can survive within a prethermal plateau with a sufficiently long life time $\sim \mathcal{O}(e^{\omega/J})$  under the evolution governed by the periodically driven Hamiltonian.(\ref{eq:Ham}) with fast driving before the system is finally heated up towards an infinite temperature state\cite{Supplementary}.  

We perform  DMRG simulation on the 2D system with a cylindrical geometry ($L_y \ll L_x$), which satisfies the periodic (open) boundary condition along y (x) direction. In the following,  we choose a cylindrical lattice  with a fixed length $L_x$ but various width $L_y$. The convergence of our results on the DMRG bond  dimension $D$ has been checked numerically\cite{Supplementary}. To characterize the ground state of $H^1$, we calculate both the single-particle and pair correlation functions, which are defined as:
\begin{eqnarray}
\nonumber C(r)&=& \langle a^\dag_{i_x,i_y} a_{i_x+r,i_y}\rangle \\
\nonumber D_{yy}(r)&=& \langle a^\dag_{i_x,i_y} a^\dag_{i_x,i_y+1}  a_{i_x+r,i_y}a_{i_x+r,i_y+1}\rangle\\
\nonumber D_{yx}(r)&=& \langle a^\dag_{i_x,i_y} a^\dag_{i_x,i_y+1}  a_{i_x+r,i_y}a_{i_x+r+1,i_y}\rangle \\
D_{xx}(r)&=& \langle a^\dag_{i_x,i_y} a^\dag_{i_x+1,i_y}  a_{i_x+r,i_y}a_{i_x+r+1,i_y}\rangle \label{eq:correlation}
\end{eqnarray} 
where $C(r)$ is the single particle correlation function between the sites $(i_x,i_y)$ and $(i_x+r,i_y)$, and $D_{yy}(r)$ ($D_{yx}(r)$) is the pair correlation function between a  vertical bond $[(i_x,i_y),(i_x,i_y+1)]$ and another vertical (horizontal) bond $[(i_x+r,i_y),(i_x+r,i_y+1)]$  ($[(i_x+r,i_y),(i_x+r+1,i_y)]$), and $D_{xx}(r)$  is the pair correlation function between two horizontal bonds alone the same line. 

We first focus on the low-density case ({\it e.g.}  1/8 filling  with $N_A=N_B=\frac 1{16}L_x L_y$ where $N_A$ ($N_B$) are the total number of bosons on sublattice A (B)).  To avoid the boundary effect along x direction, we choose the reference point $(i_x,i_y)$  as $(L_x/4,L_y/2)$ and $1\le r \le L_x/2$ in Eq.(\ref{eq:correlation}). As shown in Fig.\ref{fig:fig4} (a), the single particle correlations $C(r)$ decay exponentially with $r$, indicates the absence of  single-particle BEC. In contrast, Fig.\ref{fig:fig4} (b) suggests that the pair correlations $D_{yy}(r)$ decay algebraically in distance ($D_{yy}(r)\sim r^{-\eta}$), indicating a quasi-long-range order in such a quasi-1D system. By comparing the results with different $L_y$, one can find that the power exponent $\eta<2$ and decreases with increasing width $L_y$ (see the inset of Fig.\ref{fig:fig4} b), which indicates that the quasi-long-range order in the  cylindrical lattice  could evolve to a true long-range order in a 2D lattice with $L_y\rightarrow \infty$.  This results suggest an exotic condensate where although single-particle BEC is depleted, the bosons can condensate in pairs\cite{Pashitskii2002,Volovik2002,Bendjama2005,Schmidt2006} 


Different from the pair BECs studied before, the pairing order parameter in our model exhibits an unconventional symmetry, which can be numerically verified by comparing the paring correlations $D_{yy}(r)$ and  $D_{yx}(r)$. Our numerical results show that the former is real, while the latter is pure imaginary (see Fig.\ref{fig:fig4} (b) and (c)), indicates a $\pi/2$ phase difference between the pairing order parameters along the horizontal and vertial bonds. This $\pi/2$ phase difference can be understood as following: if we consider a pair of adjacent bosons as a new boson defined on the bond, the Hamiltonian $H^1$ is actually a  hopping of this new boson between horizontal and vertical bonds, while the $i$ factor in front of the hopping suggest that the kinetic energy is minimized only when the condensates on the vertical and horizontal bonds have the phase difference $\pi/2$.  

Despite the similarity, the $\pi/2$ phase difference of the pairing order parameters in the horizontal and vertial bonds does NOT indicates a $p_x+ ip_y$ wave symmetry, because a bosonic pairing wavefunction must possess even parity, a key distinction from the odd-parity wavefunction of p-wave pairing. To verify this even parity numerically, we can calculate the pair correlation along the horizontal bond  $D_{xx}(r)$.  As shown in  Fig.\ref{fig:fig4} (d),  $D_{xx}(r)$ also decays algebraically with $r$ but does not change sign, which implies that the phase of the pairing order parameter is the same for two adjacent horizontal bonds. This behavior contrasts with the $\pi$ phase difference characteristic of the $p_x+ip_y$ pairing state. The phase structure of the pairing condensate in our model is presented in Fig.\ref{fig:fig1} (b), and a straightforward calculation identifies it as $s+id_{x^2-y^2}$ wave pairing, which has was discussed in the context of high-Tc superconductor\cite{Kotliar1988b,Lee2009}. 

The $s+id$ pairing state, while breaking time-reversal symmetry, is nonchiral in contrast to the $p_x+ip_y$ state and thus sustains no chiral edge current. However, it is known that a supercurrent can be induced by spatial inhomogeneity, {\it e.g.} a non magnetic impurity\cite{Lee2009}. To test this within our bosonic model, we place a non-magnetic impurity at the origin point $(L_x/2,L_y/2)$  by locally shifting its chemical potential to $\mu_c$, and calculate the induced currents. As shown in Fig.\ref{fig:fig4} (e), localized currents emerge around the non-magnetic impurity and decay exponentially with distance.Because the $s+id_{x^2-y^2}$ pairing state is nonchiral, the emergent current does not form a circulating loop around the impurity. Instead, it develops a two-in two-out pattern [9] that respects the reflection symmetries about the $x$ and $y$ axes of the Hamiltonian.(\ref{eq:Ham}).

In the dilute limit, the boson pairs are well separated in distance, and an unpaired boson cannot move thus there is no single-particle BEC. In contrast, at high density, the boson pairs are sufficiently close and entangle with each other, and a boson can hop from one pair to another thus can propagate in this background of dense boson pairs. As a consequence, we expect the single-particle BEC will be recovered at high density, and the ground state is highly entangled, which makes the DMRG simulations more challenging  compared to the 1/8 case.   To verify this point, we calculate the correlation functions at half-filling case with $N_A=N_B= \frac{1}{4}L_x L_y$. As shown in Fig.\ref{fig:fig4} (f), both $C(r)$ and $D_{yy}(r)$ decays algebraically with $r$, thus we expect the ground state of Eq.(\ref{eq:Ham1}) at half-filling in a 2D lattice is a BEC phase with both single-particle and pair condensate.                     

 {\it Experimental realizations --} The Hamiltonian in Eq.(\ref{eq:Ham}) is ready to  be realized in transmons superconducting quantum circuit, where a hard-core boson can be mapped to a qubit, and their hopping amplitude can be adjusted via a coupler placed in each bond of the square lattice\cite{Yan2018b,Shi2023,Li2023}.  By periodically adjusting the coupler's frequency, we can tune the effective coupling between the adjacent qubits from positive to negative values\cite{Liu2026}, which in turn allows for the realization of the sine and cosine terms in Eq.~\eqref{eq:Ham}\cite{Supplementary}.  To detect the pairing between the bosons, we can prepare an initial state with two adjacent photons only, and let the system evolve under the periodically driven Hamiltonian.(\ref{eq:Ham}). At the end of the evolution, we can perform the measurement  to locate the position of the photons. For each experimental run, we could obtain the relative distance between the two photons, which is averaged by repeating the experiment to obtain the average distance $r(t)$. As we analyzed above, for a fast driving, we expect that $r(t)$ barely changes, while it rapidly increases for a slow driving.


{\it Conclusion and outlook --} In summary, we proposed a periodically driven protocol that enables us to implement a Floquet Hamiltonian with multisite interaction and realize an unconventional bosonic pairing condensate without single-particle BEC. It reveals new opportunities to explore the unconventional paring states in the context of non-equilibrium quantum many-body physics. The existence of extensively degenerate dark states in the FLoquet Hamiltonian $\hat{H}^1$ imposes strong kinetic constraint on the system dynamics, wherein non-ergodic dynamical phenomena is expected. For example, one may wonder whether there exits disorder-free localization, which was proposed in strongly correlated systems with multisite interaction and kinetic constraint\cite{Smith2017,Brenes2018,Karpov2021}. In addition, whether the Hilbert space fragmentation induced by the dark states in our model leads to the quantum scar states\cite{Heller1984,Turner2018} is another interesting question worthy of further studies.

{\it Acknowledgments}.--- We acknowledge helpful discussion with Congjun Wu, Heng Fan and Yunhao Shi. ZC is supported by the National Key Research and Development Program of China (2024YFA1408303), Natural Science Foundation of China (No.12525407),  Shanghai Municipal Science and Technology Major Project (Grant No.2019SHZDZX01), Shanghai Science and Technology Innovation Action Plan(Grant No. 24Z510205936). MQ acknowledges the support from the National Key Research and Development Program of MOST of China (2022YFA1405400), the National Natural Science Foundation of China (Grant No. 12274290 and No. 12522406), and the Innovation Program for Quantum Science and Technology (2021ZD0301902).

\bibliography{real}

\end{document}